\documentclass[preprintnumbers,floats,twocolumn,prd,aps]{revtex4}
\usepackage{amssymb,amsmath,epsfig}

\begin{document}

\title{The naked singularity in the global structure of critical collapse spacetimes}
\author{Andrei V. Frolov}\email{frolov@cita.utoronto.ca}
\author{Ue-Li Pen}\email{pen@cita.utoronto.ca}
\affiliation{
  CITA, University of Toronto\\
  Toronto, ON, Canada, M5S 3H8
}
\date{\today}

\begin{abstract}
  We examine the global structure of scalar field critical collapse
  spacetimes using a characteristic double-null code. It can integrate
  past the horizon without any coordinate problems, due to the careful
  choice of constraint equations used in the evolution. The limiting
  sequence of sub- and supercritical spacetimes presents an apparent
  paradox in the expected Penrose diagrams, which we address in this
  paper. We argue that the limiting spacetime converges pointwise to a
  unique limit for all $r>0$, but not uniformly. The $r=0$ line is
  different in the two limits. We interpret that the two different
  Penrose diagrams differ by a discontinuous gauge transformation. We
  conclude that the limiting spacetime possesses a singular event, with
  a future removable naked singularity.
\end{abstract}

\pacs{}
\keywords{}
\preprint{CITA-2003-28}
\maketitle

\section{Introduction}

The critical collapse of scalar fields gives rise to a new class of
thought experiments in general relativity
\cite{Choptuik:1993jv,Gundlach:2002sx}. It has been suggested that a
weak singularity may be visible to a distant observer during the
collapse of a scalar field \cite{Israel:1986,Burko:1998az}. With the
presence of such potentially problematic phenomena, one can ask to what
degree cosmic censorship is violated. The spirit of the cosmic
censorship conjecture is that the evolution of spacetimes as seen by
distant observers in asymptotically flat regions is determined uniquely
by the Einstein equations. If visible infinite curvature arises in the
evolution from regular initial data, one might have cause for concern
about the classical completeness of Einstein's theory of gravity. The
critical collapse is a fine-tuned limit, and ``strong cosmic
censorship'' has been formulated to exclude such rare cases with zero
measure in the space of initial conditions. It is nevertheless
instructive to understand the nature of this singularity, and how such
a limit can be taken.

Recently, Martin-Garcia and Gundlach \cite{Martin-Garcia:2003gz} have
numerically constructed a self-consistent discretely self-similar
spacetime which they argued to be related to the evolutionary critical
collapse solutions. They proposed the existence of a future Cauchy
horizon emanating from the critical collapse event, on which new data
must be specified. The authors proceeded to find a unique way of
specifying this data which can result in a regular future spacetime.

In this paper, we consider the problem from a different perspective.
Instead of searching for the critical solution of Einstein equations by
first taking the limit of discrete self-similarity, we study the
constructive sequence of non-critical global spacetimes. We then search
for a unique limit as we approach criticality. Posed in such a way, the
existence of a Cauchy horizon would be very unexpected: each spacetime
in the limit sequence has no Cauchy horizon, so why would it form in the
limit? To study the problem, we developed a characteristic code which
can track a scalar field collapse interior to the horizon. In Section
\ref{sec:paradox}, we outline a conceptual paradox in the search for a
global critical spacetime structure. In Section \ref{sec:spherical}, we
derive the scalar field equations of motion with spherical symmetry. We
proceed to solve these equations numerically in Section \ref{sec:code}.
The numerical results are presented in Section \ref{sec:results}, where
we explain our proposed solution to the apparent paradox.

\section{The apparent paradox}\label{sec:paradox}

\begin{figure}
  \centerline{\epsfig{file=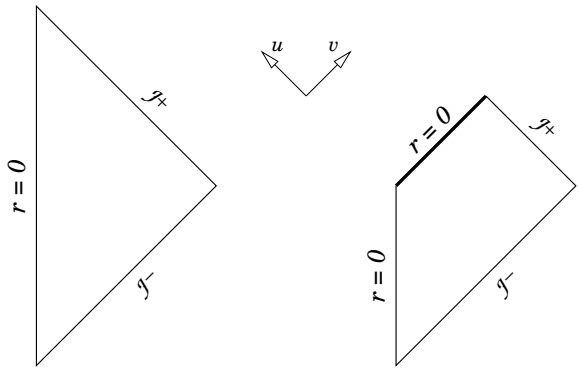, width=3in}}
  \caption{
    Two possible global structures of a critical collapse spacetime.
    The left is the diagram expected taking a limit of subcritical
    collapses, while the right diagram is a zero mass black hole
    resulting from the limit sequence of supercritical collapse.
  }
  \label{fig:comp}
\end{figure}

Consider a collapsing scalar field with amplitude characterized by a
parameter $p$. For small amplitudes $p<p_*$, the field collapses and
re-expands. For large amplitudes $p>p_*$, a black hole forms. An
interesting question is to examine the behavior when $p=p_*$. One can
consider two limits, one from below and one from above. If we consider
a sequence of subcritical spacetimes with collapsing scalar fields as
they approach $p_*$ from below, we might expect the global structure of
the resulting spacetime to be Minkowski, i.e. a triangle shown in the
left panel of Fig.~\ref{fig:comp}. If we consider the limit of a
sequence of supercritical spacetimes from above, we expect each stage
to have a global Schwarzschild structure. As the parameter decreases,
the mass of the resulting black hole approaches zero, and one expects
the limiting spacetime to resemble a Schwarzschild spacetime with zero
mass, shown in the right panel of Fig.~\ref{fig:comp}. In this limit,
$r=0$ coincides with the horizon and becomes null (and infinitely
redshifted). The global structure looks quadrangular, quite unlike the
argued spacetime in the other limit.

This suggests several possible interpretations. Perhaps the two limits are
different, and the limiting spacetime depends on the direction from
which the limit was taken. Or one of the limits is only an incomplete
description of spacetime, and might be extensible to the same global
structure. Or perhaps the limit is not convergent from either direction,
and oscillates in such a way that more data must be specified on a
spontaneously formed Cauchy horizon \cite{Martin-Garcia:2003gz}. Our
study suggests a slightly different physical interpretation of the global
structure.

Critical spacetimes are only known as numerical solutions, which makes
questions about global structure hard to answer. It is most easily
studied in characteristic coordinates which follow light ray propagation
\cite{Garfinkle:1995jb}. In the subsequent section we will describe our
formulation and implementation of the numerical procedures.

\section{Spherically Symmetric Scalar Field Collapse}\label{sec:spherical}

The spherically symmetric $(n+2)$-dimensional spacetime metric can be
written as
\begin{equation}\label{metric}
  ds^2 = e^{-2\sigma} d\vec{x}^2 + e^{2\mu} d\Omega^2,
\end{equation}
where $d\Omega^2$ is the metric of a unit $n$-dimensional sphere with
curvature $K=1$, and the two-manifold metric $d\gamma^2 = e^{-2\sigma}
d\vec{x}^2$ is conformally flat. The dynamics of the scalar field
collapse are described by the reduced action
\begin{eqnarray}\label{action}
  S &\propto& \int d^2 x\, e^{n\mu} \Big\{
    - 2 (\nabla\phi)^2
    - 2n(\nabla\sigma \cdot \nabla\mu) \\
  &&\hspace{5em}
    + n(n-1) \left[(\nabla\mu)^2 + K e^{-2(\sigma+\mu)}\right]
  \Big\},\nonumber
\end{eqnarray}
where integration and differential operators are with respect to the
flat two-metric $d\vec{x}^2$. Variation of the above action with
respect to the fields $\phi$, $\mu$ and $\sigma$ gives equations of
motion
\begin{subequations}\label{eom}
\begin{eqnarray}
  &\Box\phi + n (\nabla\mu \cdot \nabla\phi) = 0,& \\
  &\Box\mu + n (\nabla\mu)^2 - (n-1)K e^{-2(\sigma+\mu)} = 0,& \\
  &\Box\sigma - \frac{n}{2} \left\{\Box\mu + (\nabla\mu)^2\right\} - (\nabla\phi)^2 = 0,&
\end{eqnarray}
\end{subequations}
while the two constraint equations are recovered by variation with
respect to the (flat) metric
\begin{equation}\label{constraint}
  \text{Traceless} \left\{
    \mu_{;ab} + \mu_{,a}\mu_{,b} + 2 \mu_{(,a}\sigma_{,b)}
    + {\textstyle \frac{2}{n}} \phi_{,a}\phi_{,b}
  \right\} = 0.
\end{equation}
For the particular case of spherically symmetric scalar field collapse
in four dimensions ($n=2$), the equations of motion (\ref{eom}) can be
simplified by introducing auxiliary field variables $r = e^\mu$ and
$\varphi = r\phi$:
\begin{subequations}\label{evolve}
\begin{eqnarray}
  &\Box(r^2) = 2 e^{-2\sigma},& \\
  &\Box\varphi = \frac{\Box r}{r}\, \varphi,& \\
  &\Box\sigma = \frac{\Box r}{r} + (\nabla\phi)^2.&
\end{eqnarray}
\end{subequations}
This form of the dynamical equations is better suited for numerical
evolution.

\section{Characteristic Code}\label{sec:code}

We discretize and evolve the collapsing scalar field spacetime in
double-null coordinates $d\vec{x}^2 = -2\, du\, dv$, where the radial
characteristics of the wave equations are made explicit: the outgoing
characteristic propagates at constant $u$ in the direction of
increasing $v$, while the incoming characteristic propagates at
constant $v$ in the direction of increasing $u$. This approach has a
number of advantages over some of the more traditional spacetime
slicings.

\begin{figure*}[t!]
  {\hfill
  \epsfig{file=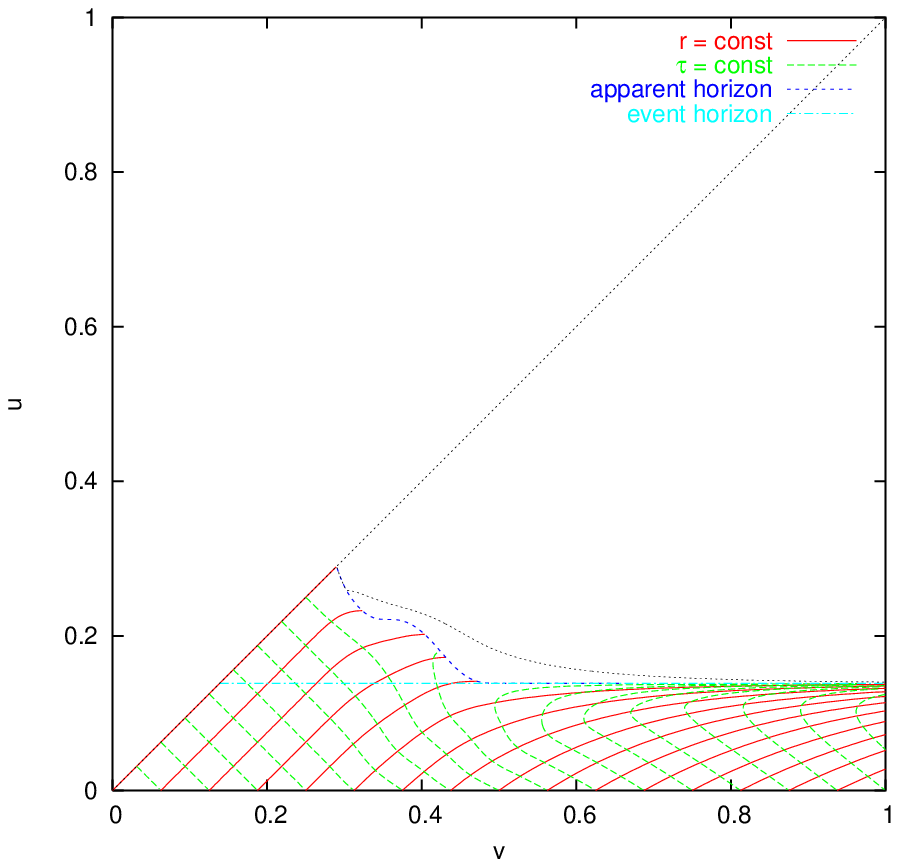, height=3in}
  \hfill
  \epsfig{file=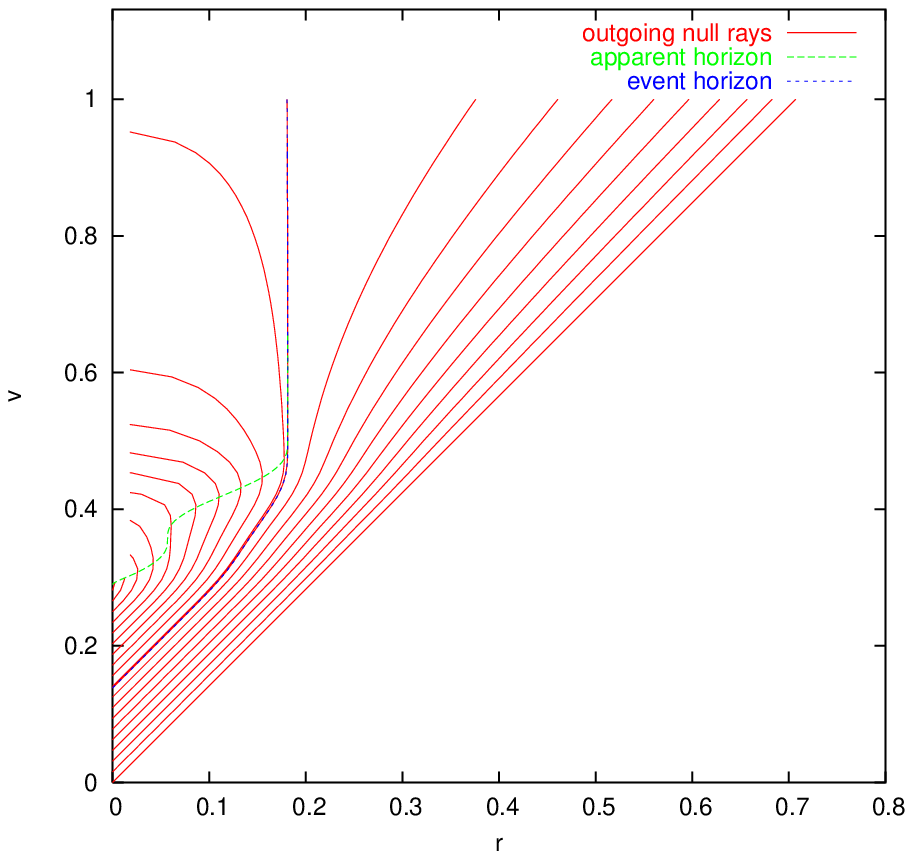, height=3in}
  \hfill}
  \caption{
    Penrose (left) and Kruskal (right) style diagrams of the spacetime
    with a large black hole produced in the collapse of a sine-squared
    scalar field wavepacket (\protect\ref{sin2}). The grid lines on the
    Penrose diagram show observers at constant radius and the proper
    time according to their clocks (note that the plot is rotated
    $45^\circ$ compared to Fig.~\protect\ref{fig:comp}; the diagonal
    line $u=v$ is the origin $r=0$). The Kruskal diagram shows the
    trajectories of outgoing null rays as emitted from the center of
    spacetime at subsequent moments of time. The code sees both real
    and apparent horizons, as well as the formation of the spacelike
    singularity (small region around which is excised). Note that the
    apparent and event horizons are not the same, as the spacetime is
    not static.
  }
  \label{fig:bh}
\end{figure*}

The characteristic code only propagates information along
characteristics at a numerical speed equal to the true characteristic
speed. The numerical domain of dependence is no larger than the
physical domain of dependence, and the code still maintains full
(second order) accuracy. Horizons are not particularly special as far
as ingoing null characteristics are concerned. This allows us to follow
collapse all the way to the singularity. Even as an outgoing
characteristic hits a singularity and the floating point numbers
denormalize, this does not affect any of the other characteristics,
which can be integrated to fill the whole maximally extended spacetime
determined by the initial data. To illustrate this point, in
Fig.~\ref{fig:bh} we present Penrose and Kruskal diagrams of a
spacetime with a large black hole formed in the collapse of the scalar
field wavepacket.

The two-dimensional metric $d\gamma^2$ has a residual gauge freedom under
redefinition of the null coordinates $u \mapsto U(u)$, $v \mapsto V(v)$.
These two free functions (of a single variable) are used to define the
coordinate $v$ on the initial slice and to map the central point $r=0$
to a straight line in the $(u,v)$ plane:
\begin{equation}\label{gauge}
  v|_{u=0} = \sqrt{2}\, r, \hspace{1em}
  u|_{r=0} = v.
\end{equation}
The second gauge condition is particularly convenient, since it places
the central point at a known location on the grid when discretizing.

The initial conditions are specified on a surface of constant $u$ by
giving a scalar field profile $\phi(v)$. Together with the gauge choice
(\ref{gauge}), this determines the rest of the variables. In
particular, $\sigma(v)$ is obtained by integrating the outgoing ($vv$)
constraint equation (\ref{constraint})
\begin{equation}\label{ic}
  \sigma|_{u=0} = -\frac{1}{2} \int \phi_{,v}^2 v dv.
\end{equation}
The integration is implemented as a fourth order Runge-Kutta algorithm
with a fixed step. Although strictly speaking it is not necessary, as
the evolution code is second order, it is no more complicated than a
second order integrator would be.

Although the constraint equations follow from the dynamical equations,
their use might be required for stability \cite{Gundlach:1997rk}.
Using the outgoing constraint equation for evolution (rather than just
for initial conditions) is not a good idea, however. It becomes
degenerate on the apparent horizon, where the outgoing light rays
become (marginally) trapped $r_{,v}=0$. The incoming constraint
equation is perfectly fine, though, and could be followed all the way
to the center of the spacetime, whether it is singular or not. By
introducing an auxiliary field variable $\tau = \sigma + \frac{1}{2}
\ln\left(-\sqrt{2}\, r_{,u}\right)$, the incoming constraint equation
(\ref{constraint}) can be written in a form that is very simple to
integrate
\begin{equation}
  \tau_{,u} = -\frac{1}{2} \frac{\phi_{,u}^2}{(\ln r)_{,u}}.
\end{equation}
To integrate the incoming constraint, one would need to know the values
of $\tau$ on the initial slice $u=\text{const}$. These can be found by
integration of the equation
\begin{equation}
  \tau_{,v} = -\frac{1}{2} \phi_{,v}^2 v - \frac{1}{2v} \left[1 - e^{-2\tau}\right],
\end{equation}
which is obtained by combining the constraints with the evolution
equation (\ref{evolve}a) to solve for $r_{,u}$ on the initial slice.

\begin{figure}[h]
  \centerline{\epsfig{file=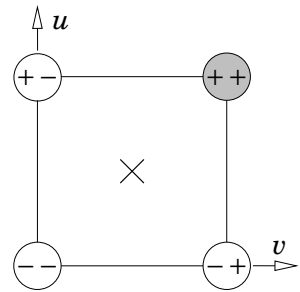, width=1.5in}}
  \caption{Discretization on a $(u,v)$ grid.}
  \label{fig:discrete}
\end{figure}

Having discussed our gauge choice, initial conditions and constraint
equations, we now come to the discretization of the evolution
equations. The covariant differential operators in the evolution
equations (\ref{evolve}) are with respect to a flat metric, so in null
coordinates they are written simply as
\begin{equation}
  \Box x = -2 \partial_u\partial_v x, \hspace{1em}
  (\nabla x)^2 = -2 (\partial_u x) (\partial_v x),
\end{equation}
where $x$ denotes any one of the three dynamical variables in equations
(\ref{evolve}). We discretize by finite differencing on a rectilinear
$(u,v)$ grid with equal spacing $du=dv=\Delta$, which to second order
accuracy gives
\begin{eqnarray}
  (\Box x)_\times &=& -\frac{2}{\Delta^2} \left[x_{++} + x_{--} - x_{+-} - x_{-+}\right], \\
  (\nabla x)^2_\times &=& -\frac{1}{2 \Delta^2} \left[(x_{++} - x_{--})^2 - (x_{-+} - x_{+-})^2\right], \nonumber
\end{eqnarray}
where the differential operators are evaluated at the center of a grid cell
(see Fig.~\ref{fig:discrete}).
The code takes a step by using discretized evolution equations to find
the values of the fields at the $(++)$ grid point (shaded node in
Fig.~\ref{fig:discrete}). The only non-trivial operation involved in
this is finding $r_{++}$ accurate to second order, which is done by
solving equation (\ref{evolve}a) discretized in the following fashion:
\begin{eqnarray}
  && r_{++}^2 + r_{--}^2 - r_{+-}^2 - r_{-+}^2 = \\
  && \hspace{1em}
  \frac{\Delta}{\sqrt{2}} \exp(-\tau_{+-}-\tau_{-+}) (r_{++} + r_{+-} - r_{-+} - r_{--}).
  \nonumber
\end{eqnarray}
The rest is then straightforward, as the right hand sides of equations
(\ref{evolve}b,c) become known.

As the code advances to the next slice of constant $u$, the very first
point on the grid ($u=v$) is the center of the spacetime ($r=0$) and
has to be treated specially. The asymptotic form of the evolution
equations at $r=0$
\begin{equation}
  \textstyle
  \nabla r \cdot \nabla\phi = 0, \hspace{1em}
  \sigma = -\frac{1}{2} \ln(\nabla r)^2
\end{equation}
is used to calculate the values of the fields in the center.

\section{Results}\label{sec:results}

We tested the code on the collapse of a scalar field with various
initial conditions. In particular, we used pulse
\begin{equation}\label{sin2}
  \phi(v) = \left\{
	\begin{array}{cc}
	  p \sin^2 4\pi v, & \frac{1}{4} < v < \frac{1}{2} \\
	  0 & \text{otherwise} \\
	\end{array}
  \right.
\end{equation}
and kink
\begin{equation}\label{tanh}
  \phi(v) = \left\{
	\begin{array}{cc}
	  p + p \tanh \left[5 \tan\pi(\frac{4}{3}v - \frac{1}{2})\right], & v < \frac{3}{4} \smallskip\\
	  2p, & v \ge \frac{3}{4} \\
	\end{array}
  \right.
\end{equation}
field profiles in our simulations. Both profiles are fairly smooth
functions with field energy density having compact support on the
initial slice. This avoids the interference of long-range tails in the
initial data on the late-time evolution.

\begin{figure}
  \centerline{\epsfig{file=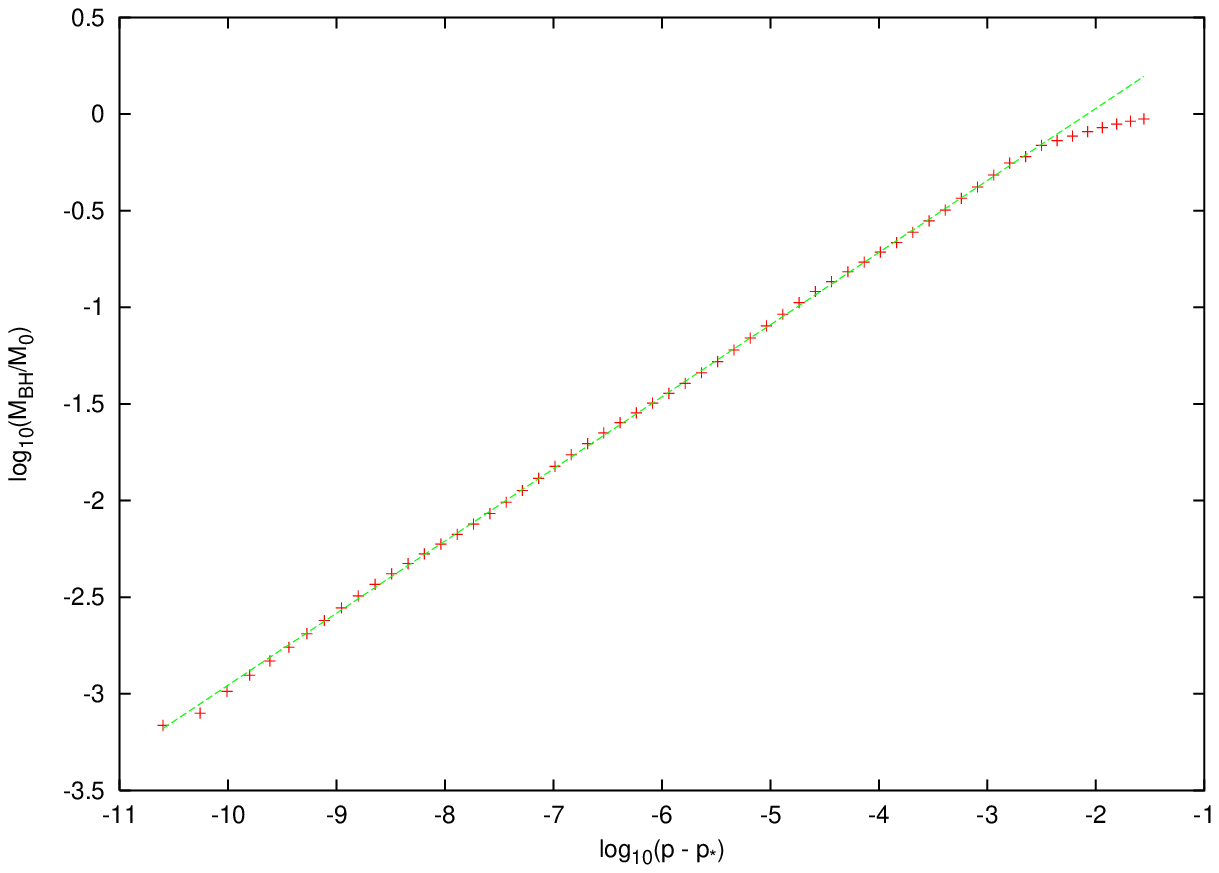, width=\columnwidth}}
  \centerline{\epsfig{file=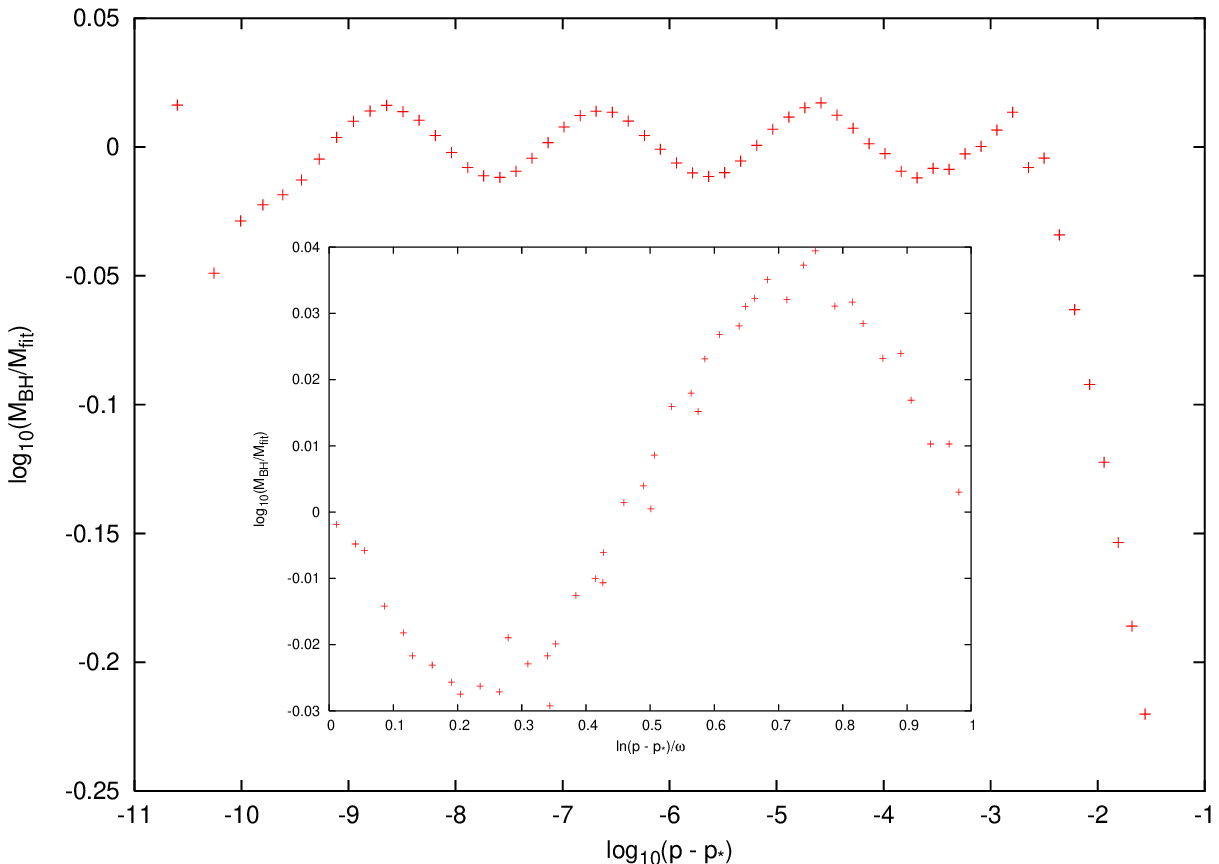, width=\columnwidth}}
  \caption{
    Black hole mass scaling for single kink wave, given by equation
    (\protect\ref{tanh}). The upper panel shows a power law fit to the
    numerical data with critical exponent $\gamma \approx 0.373$ which
    holds over three orders of magnitude in black hole masses,
    corresponding to nine orders of magnitude in the tuning parameter.
    The lower panel shows the residual after the power law fit,
    demonstrating periodic fine structure due to discrete
    self-similarity of the collapse. Its period $\omega \approx 4.63$
    corresponds to the value of the echoing parameter $\Delta \approx
    3.45$. The lower panel inset shows the residual mapped to a single
    period (outlier points removed), illustrating that the numerical
    data is in good agreement with a periodic modulation of the black
    hole mass.
  }
  \label{fig:scaling}
\end{figure}

All runs shown in this paper used 65536 uniformly spaced grid points.
The critical scaling of black hole mass is shown in Fig.~\ref{fig:scaling}.
It agrees well with the literature \cite{Choptuik:1993jv,Hod:1997az}.
Our achievable dynamic range shows scaling over three orders of
magnitude in black hole mass without the use of adaptive mesh
refinement. This is sufficient for our study. Further improvements in
dynamic range can be sought after using adaptive mesh refinement
techniques or adopting an initial gauge which places ingoing null rays
on a grid more densely near the collapse point \cite{Garfinkle:1995jb}.
As the grid resolution is increased, the double arithmetic precision
and round-off errors become the main obstacles in the quest for higher
dynamical range.

\begin{figure}
  \centerline{\epsfig{file=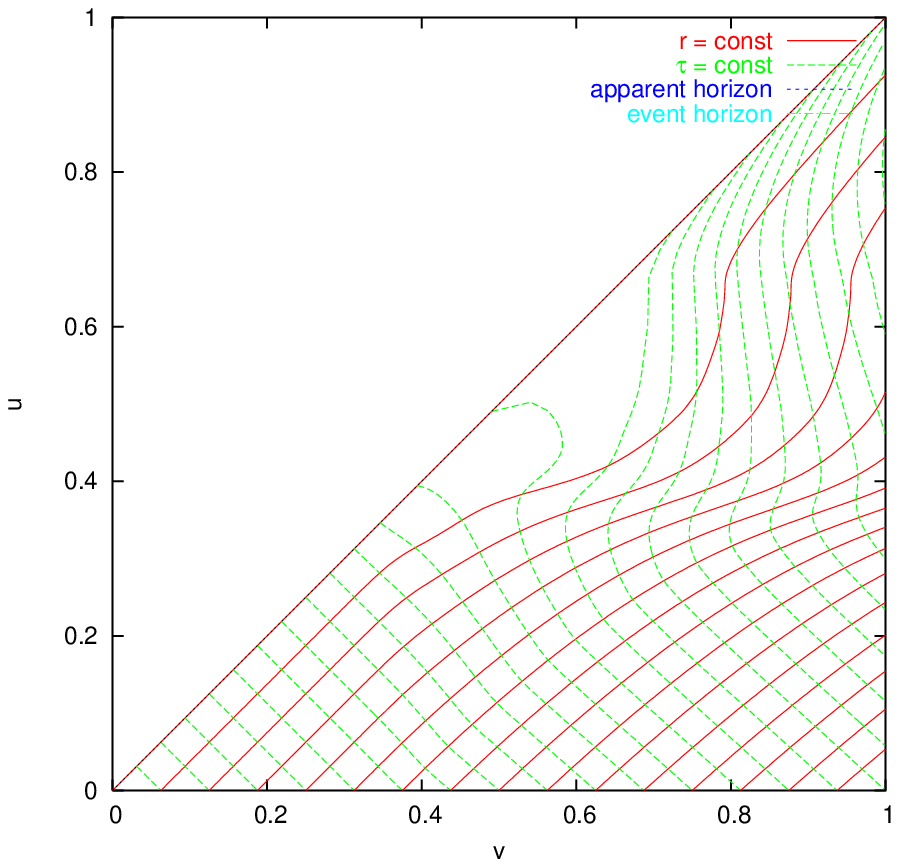, height=2.75in}}
  \vspace{-0.5cm}
  \caption{Penrose diagram for subcritical collapse.}
  \label{fig:subcrit}
  \medskip
  \centerline{\epsfig{file=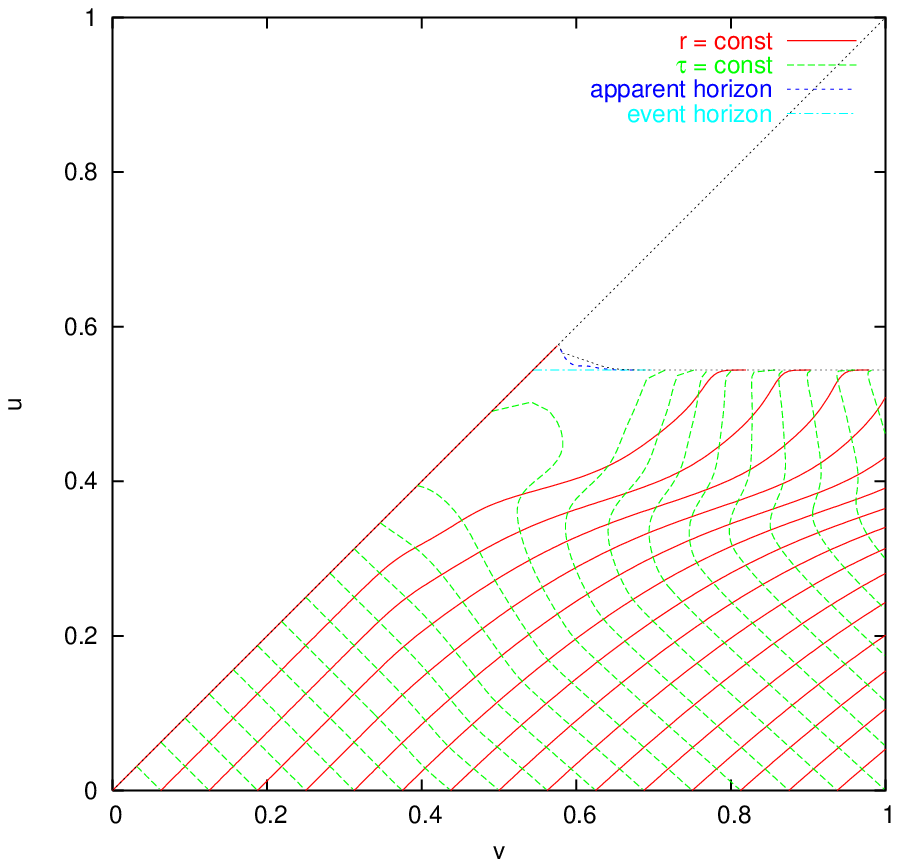, height=2.75in}}
  \vspace{-0.5cm}
  \caption{Penrose diagram for supercritical collapse
    with $M_{\text{BH}}/M_0 \approx 0.046$.}
  \label{fig:supercrit}
  \medskip
  \centerline{\epsfig{file=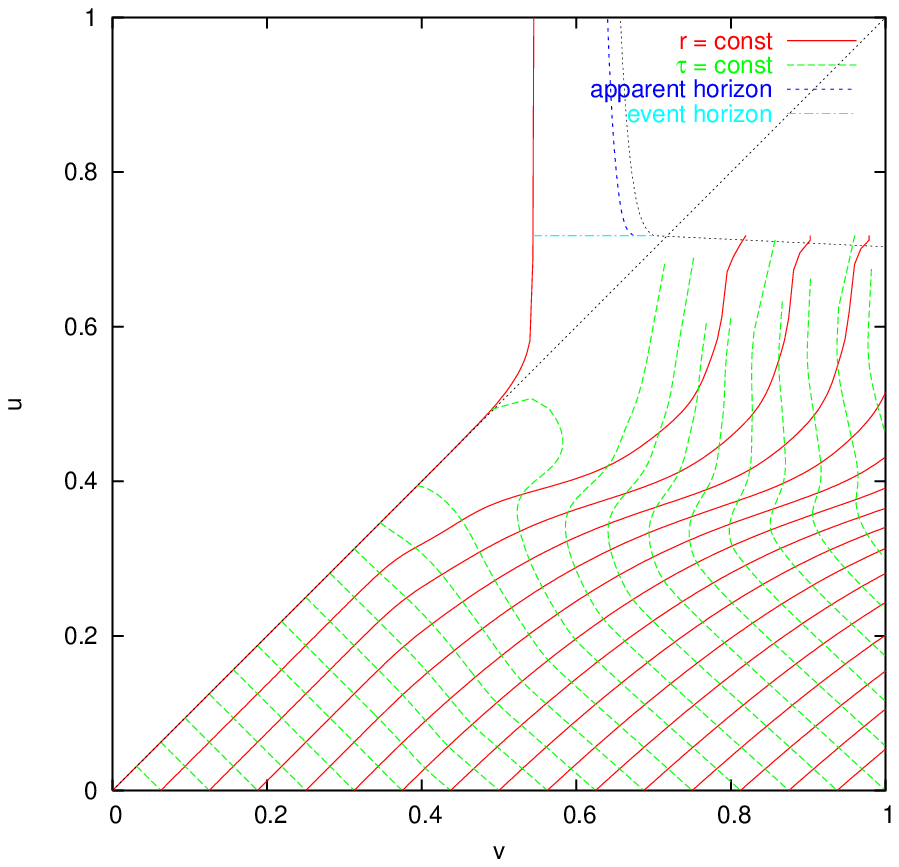, height=2.75in}}
  \vspace{-0.5cm}
  \caption{
    Re-gauged supercritical collapse of Fig.~\protect\ref{fig:supercrit}.
    The re-gauging was done at the postprocessing stage from the same
    simulation run. One sees the clear similarity with the subcritical
    collapse shown in Fig.~\protect\ref{fig:subcrit}.
  }
  \label{fig:stretch1}
\end{figure}

Fig.~\ref{fig:subcrit} shows the Penrose diagram of a subcritical
spacetime just below the threshold of black hole formation. The lines
of constant radius $r$ are drawn as solid curves, and the proper time
along such lines are drawn as dashed curves. Most of the scalar field
mass is shed before $u \sim 0.6$, but the field oscillating on ever
smaller scales creates near-singular curvature in the center of
spacetime.

When we examine the Penrose diagram of a supercritical collapse as
shown in Fig.~\ref{fig:supercrit}, we see the formation of a horizon
near $u\approx 0.55$. The code evolves well inside the event horizon,
and we can identify the apparent horizon in the interior of the black
hole. The lines of constant $r$ become spacelike beyond the apparent
horizon. All the larger radii outside the black hole pile up at the
horizon in this diagram.

We can pose the question if the spacetime described by
Fig.~\ref{fig:supercrit} could possibly be the same as
Fig.~\ref{fig:subcrit}. Our original gauge choice from equation
(\ref{gauge}) fixes the $r=0$ line to coincide with $u=v$. It does not
have to be this way; the null coordinates leave the possibility for a
global gauge change $u \mapsto U(u)$. One can try to ``un-pile'' the
lines of constant $r$ observers for the supercritical collapse
spacetime in Fig.~\ref{fig:supercrit} by defining a gauge change $U(u)$
such that the second $r$ curve coincides with that in a subcritical
collapse spacetime of Fig.~\ref{fig:subcrit}. The result is shown in
Fig.~\ref{fig:stretch1}.

This stretching is rather sudden and throws the $r=0$ line into an
almost null direction, but all the other finite $r$'s appear to
coincide. This is non-trivial, since the residual gauge freedom is a
single function of one variable, which we used to fix one value $r$. If
the spacetimes are equivalent, the other $r$ and $t$ should fall into
place, and similarly if the spacetimes are different they should
diverge. We indeed see that the proper time and outer radii move into
place, as would be expected in a convergent spacetime. It is also clear
that the $r=0$ line is not convergent, and the limiting spacetime
experiences infinite curvature at that one point. To address the future
of this singular event, we can look at the spacetime of the limiting
sequence of super- and subcritical collapses.

We now discuss the global structure of the critical spacetime. Our
first question is the nature of the $r=0$ line in the course of the
singular collapse event. An observer moving at some finite radius $r$
sees the collapse of a scalar field to a point, and a rebound. This is
true for both the supercritical and subcritical collapse, since even
when a black hole forms, the majority of the initial scalar field
energy escapes, and the black hole only contains an ever smaller
fraction of the initial mass as the parameter is tuned to criticality.
Long after the field rebounds, the observer can measure the
gravitational redshift to adjacent interior radii, and concludes that
there is no redshift for most radii. In the supercritical scenario,
there are significant redshifts at scales $r\lesssim M_\text{BH}$. But
for any fixed radius observer, as one takes the limit of $p \rightarrow
p^*$, the sphere of influence of the ever diminishing black hole mass
shrinks to zero, and the spacetime converges pointwise to Minkowski at
$r>0$. The convergence is not uniform, since for any given redshift
difference $\epsilon$ between the observer and a fiducial interior
radius, one can construct a radius $r\sim M/\epsilon$ inside which the
redshift is larger than $\epsilon$. In the two limits $p\rightarrow
p^*_+$ and $p\rightarrow p^*_-$, the spacetimes converge pointwise
everywhere except for the line $r=0$.

A simple analogy is a singular weak field star. Consider a sequence of
spacetimes with a single star of mass $M_*$ and radius
$r_*=M_*/\epsilon$. The Newtonian potential outside the star is
$V(r)=-M_*/r$ for $r>M_*/\epsilon$. To simplify the argument, we make
the potential continuous and constant interior to its radius
$V(r)=\epsilon$. For small values of $\epsilon$, the spacetime is in
the weak field regime everywhere to order $\epsilon$. If we take a
sequence of such space times with fixed $\epsilon$ and decreasing
$M_*$, the maximal curvature $R\propto \epsilon^3/M_*^2$ increases
without bound. There appears to be an illusory naked singularity in
that limit. We can ask what the limiting spacetime looks like as $M_*
\rightarrow 0$. The obvious answer would be empty Minkowski space. The
convergence to this space is pointwise, but not uniform. In the limit,
$V_0(r)=\lim_{M_*\rightarrow 0}M_*/r$. $V_0(r)$ converges to 0
pointwise, but not uniformly, so in the limit, $V_0(r)=0$ everywhere
except for $r=0$, and it is undefined at that point. This is a
removable singularity since we can define $V_0(0)\equiv
\lim_{r\rightarrow 0} V_0(r) =0$, which is a unique and physically
acceptable solution. We suggest that the future of the singular
collapse is likewise regular everywhere, with a removable singularity
at $r=0$ in the future of the collapse event.

The apparent paradox in the Penrose diagram seen in Fig.~\ref{fig:comp}
arises from the gauge choice at $r=0$. This one line is poorly defined
in the limit, since it becomes singular. The left panel describes the
physical spacetime in the critical collapse limit, and the right one is
related by a singular gauge change.

Our analysis takes a very different approach from Martin-Garcia and
Gundlach \cite{Martin-Garcia:2003gz}. These authors solve an elliptic
equation satisfying an ansatz of discretely self-similar critical
collapse. We took a limit of a series of hyperbolic initial value
problems. Those authors found a possibility of specifying new Cauchy
data in the elliptic solution, which is never an option for our
evolutionary approach. The qualitative solution for their unique
regular extension looks similar to our critical spacetime.

\section{Conclusions}

We have implemented a characteristic code in double-null coordinates
and used it to study the global spacetime structure of a critical
collapse of a scalar field. We reproduce the standard critical behavior
and universal scaling. We compare the limiting spacetime from the
subcritical and the supercritical collapse limits. These two limits
appear qualitatively different. Based on the numerical simulations, we
conjecture that the two limits converge pointwise to the same
spacetime, but not uniformly. In particular, the $r=0$ line is not
convergent, but all other points appear to converge pointwise. The two
apparently different solutions then only differ by a gauge change. The
apparent naked singularity in the upper limit obtained by the sequence
of spacetimes with ever decreasing black hole mass becomes a removable
singularity in the limit.

We conclude that the nature of the limiting critical collapse space
time requires a careful definition of the order in which limits are
taken, since the convergence is not uniform. For any collapse parameter
$p$ within the critical value $|p-p^*|=\delta>0$, one can find
sufficiently small radii within which the supercritical and subcritical
solutions differ. Conversely, for any fixed $r>0$, one can find a value
$\delta(r)$ where the super and sub critical space times agree to some
tolerance $\epsilon$ at radius $r$. There is a unique limit, with no
new Cauchy data that emits from the singular event.

\bigskip
\section*{Acknowledgments}

We would like to thank David Garfinkle for helpful comments and Tara
Hiebert for interactions during the early stages of the project.



\end{document}